\documentclass[sigconf,10pt]{acmart}

\usepackage{listings}
\usepackage{algorithm}
\usepackage{algpseudocode}
\usepackage{comment}
\usepackage{hyperref}
\usepackage{lstautogobble}
\usepackage{listings}
\usepackage{graphicx}
\usepackage{subcaption}
\usepackage{booktabs}
\usepackage{tikz}
\usepackage{textcomp}
\usepackage{stfloats}
\usepackage{footnote}
\usepackage{enumitem}
\setlist{nolistsep}

\copyrightyear{2024}
\acmYear{2024}
\setcopyright{rightsretained}
\acmConference[ICS '24]{Proceedings of the 38th ACM International Conference on Supercomputing}{June 4--7, 2024}{Kyoto, Japan}
\acmBooktitle{Proceedings of the 38th ACM International Conference on Supercomputing (ICS '24), June 4--7, 2024, Kyoto, Japan}\acmDOI{10.1145/3650200.3656597}
\acmISBN{979-8-4007-0610-3/24/06}

\newcommand*\circled[1]{\tikz[baseline=(char.base)]{
            \node[shape=circle,draw,inner sep=1pt] (char) {\textcolor{black}{#1}};}}

\makeatletter
\DeclareUrlCommand\ULurl@@{%
  \def\UrlLeft{\uline\bgroup}%
  \def\UrlRight{\egroup}}
\def\ULurl@#1{\hyper@linkurl{\ULurl@@{#1}}{#1}}
\DeclareRobustCommand*\ULurl{\hyper@normalise\ULurl@}
\def\@copyrightpermission{\relax}
\makeatother

\newcommand{\tool}[0]{\mbox{\textsc{Snoopie}}}

\usepackage{amsmath}

\definecolor{bluekeywords}{rgb}{0.13, 0.13, 1}
\definecolor{greencomments}{rgb}{0, 0.5, 0}
\definecolor{redstrings}{rgb}{0.9, 0, 0}
\definecolor{graynumbers}{rgb}{0.5, 0.5, 0.5}

\definecolor{background_ok}{rgb}{0.95, 0.95, 0.95}

\usepackage[listings,skins]{tcolorbox}
\def\BibTeX{{\rm B\kern-.05em{\sc i\kern-.025em b}\kern-.08em
    T\kern-.1667em\lower.7ex\hbox{E}\kern-.125emX}}

\usepackage[switch]{lineno}
   
\AtBeginDocument{%
  \providecommand\BibTeX{{%
    \normalfont B\kern-0.5em{\scshape i\kern-0.25em b}\kern-0.8em\TeX}}}

\usepackage{tabularx}

\begin{document}

\title{\tool{}: A Multi-GPU Communication Profiler and Visualizer}

\author{Mohammad Kefah Taha Issa}
\authornote{These authors contributed equally}

\affiliation{%
  \institution{Ko\c{c} University}
  \city{Istanbul}
  \country{T\"{u}rkiye}}
\email{missa18@ku.edu.tr}

\author{Muhammad Aditya Sasongko}
\authornotemark[1]
\affiliation{%
  \institution{Ko\c{c} University}
  \city{Istanbul}
  \country{T\"{u}rkiye}}
\email{msasongko@ku.edu.tr}

\author{Ilyas Turimbetov}
\affiliation{%
  \institution{Ko\c{c} University}
  \city{Istanbul}
  \country{T\"{u}rkiye}}
\email{iturimbetov18@ku.edu.tr}

\author{Javid Baydamirli}
\affiliation{%
  \institution{Ko\c{c} University}
  \city{Istanbul}
  \country{T\"{u}rkiye}}
\email{jbaydamirli21@ku.edu.tr}

\author{Do\u{g}an Sa\u{g}bili}
\affiliation{%
  \institution{Ko\c{c} University}
  \city{Istanbul}
  \country{T\"{u}rkiye}}
\email{dsagbili17@ku.edu.tr}

\author{Didem Unat}
\affiliation{%
  \institution{Ko\c{c} University}
  \city{Istanbul}
  \country{T\"{u}rkiye}}
\email{dunat@ku.edu.tr}

\renewcommand{\shortauthors}{Issa et al.}
\begin{abstract}

  With data movement becoming one of the most expensive bottlenecks in computing,
   the need for profiling tools to analyze communication becomes crucial for effectively scaling multi-GPU applications.
  While existing profiling tools including first-party software by GPU vendors are robust and excel at capturing compute operations within a single GPU, 
  support for monitoring GPU-GPU data transfers and calls issued by communication libraries is currently inadequate.
  To fill these gaps, we introduce \tool{}, an instrumentation-based multi-GPU communication profiling tool built on NVBit, capable of tracking peer-to-peer transfers and GPU-centric communication library calls.
  To increase programmer productivity, \tool{} can attribute data movement to the source code line and the data objects involved. It comes with multiple visualization modes at varying granularities, from a coarse view of the data movement in the system as a whole to specific instructions and addresses. 
Our case studies  demonstrate
\tool's effectiveness in monitoring data movement, locating performance bugs in applications, and understanding concrete data transfers abstracted beneath communication libraries.
The tool is publicly available at {\color{teal} \url{https://github.com/ParCoreLab/snoopie}}. %
 
\end{abstract}



\settopmatter{printfolios=true}
\maketitle

\section{Introduction}

Application development and performance scaling on multiple GPUs is nontrivial and often require continuous profiling and debugging. 
GPU communication mechanisms are also diverse, with prior methods requiring data transfers to pass through the host CPU. However, advancements in direct peer-to-peer communication technology such as NVLink and GPUDirect \cite{gpudirectrdma} have enabled GPUs to communicate directly with their peers during data transfers, bypassing the intermediate host buffers. These transfers can be initiated by either the host or device and can involve direct access to remote memory through loads or stores. 

However, despite the advancements in data movement technologies, 
existing multi-GPU profiling tools have lagged behind, offering limited information about data movement. %
Nsight Systems  \cite{nsight-system},  a performance analysis tool developed and maintained by Nvidia, {\em vaguely} indicates the presence of device-initiated direct memory accesses as traffic between NVLink-connected GPUs but does not provide further information to end-users such as devices engaged in such communication. 
HPCToolKit \cite{hpctoolkit}, which is a suite of performance analysis tools for CPUs and GPUs, can track the amount of data moved among devices but it is restricted to cases where communication calls are initiated by the host.
Similarly, the majority of other profiling tools offer limited or no insights into data movement caused by communication libraries like NVSHMEM \cite{Nvshmem_Docs} and NCCL~\cite{nccl}. The latter is utilized by popular deep learning frameworks, including PyTorch~\cite{pytorch} and TensorFlow~\cite{tensorflow}, to facilitate multi-GPU acceleration.

Furthermore, most tools \cite{5948661, nsight-system, 10.1007/978-3-642-31476-6_7, 10.1007/978-3-030-71058-3_10} including those that do support said communication operations lack source code and object attribution, providing an incomplete picture in their visualization. Attributing communication to the source code and involved program objects can greatly aid programmers in identifying scalability bottlenecks and data movement-related bugs. As a result, a diagnostic tool to address the aforementioned gaps is needed for effective development and debugging on multi-GPU systems.

This paper presents \tool{}, a complementary profiler that is designed specifically to address the shortcomings in existing mainstream tools, with the issues mentioned above in mind. 
\tool{} provides finer-grained information on C/C++ and Python-based GPU applications by detecting host-initiated transfers and peer-to-peer communication including device-initiated direct accesses, while providing support for commonly used communication libraries. It can identify the data objects accessed by the inter-GPU memory operations and locate the source code lines where communication occurs. Lastly, it is equipped with a multi-view interactive visualization module 
that exposes various levels of granularity for users to obtain a broad overview of the system or a detailed view of a specific object or device.
In short, our main contributions are as follows:

\begin{table*}[thp]
\footnotesize
{
\begin{tabular}{lcccccc}

\toprule
{\bf Profiler   }                         & { \begin{tabular}[c]{@{}c@{}}{\bf cudaMemcpy} \\ {\bf and its family}\end{tabular}} & \begin{tabular}[c]{@{}c@{}}{\bf P2P Direct} \\ {\bf Access}\end{tabular} & \begin{tabular}[c]{@{}c@{}}{\bf Communication} \\ {\bf Library Support}\end{tabular} & \begin{tabular}[c]{@{}c@{}}{\bf Src Code Line} \\ {\bf Attribution}\end{tabular} & \begin{tabular}[c]{@{}c@{}}{\bf Data Object} \\ {\bf Attribution}\end{tabular} & {\bf Visualization }                                                           \\ \hline
\begin{tabular}[c]{@{}c@{}}Nsight  Systems~\cite{nsight-system}\end{tabular} & \checkmark                                                                                                                 & Limited                                                      & Limited                                                                      & -                                                                  & -                                                                    & timeline                                                                 \\ \hline
HPCToolkit~\cite{hpctoolkit}                                                  & \checkmark                                                                                                                 & -                                                            & -                                                                        & \checkmark                                                                    & -                                                                  & code-centric                                                             \\ \hline
ComScribe\cite{10.1007/978-3-030-71058-3_10}                           & \checkmark                                                                                                                 & \checkmark                                                          & only NCCL                                                                & -                                                                  & -                                                                    & \begin{tabular}[c]{@{}c@{}}comm  matrix\end{tabular}          \\ \hline
EZTrace\cite{5948661}                             & \checkmark                                                                                                                 & -                                                            & -                                                                        & -                                                                  & -                                                                    & \begin{tabular}[c]{@{}c@{}}timeline and \\ comm matrix\end{tabular}          \\ \hline
Extrae\cite{extrae}                              & \checkmark                                                                                                                 & -                                                            & -                                                                        & -                                                                & \checkmark                                                                    & timeline                                                                 \\ \hline
Score-P\cite{10.1007/978-3-642-31476-6_7}                             & \checkmark                                                                                                                 & -                                                            & -                                                                        & -                                                                  & -                                                                    & timeline                                                                 \\ \hline
\tool{} (ours)                            & \checkmark                                                                                                                 & \checkmark                                                          & \begin{tabular}[c]{@{}c@{}}NCCL and \\ NVSHMEM\end{tabular}              & \checkmark                                                                & \checkmark                                                                  & \begin{tabular}[c]{@{}c@{}}multi-view, code- \\ data-centric\end{tabular} \\ 
\bottomrule
\end{tabular}
}
\caption{Comparing communication profiling tools for multi-GPU systems
\vspace{-0.2in}
}
\label{tab:related-work}
\end{table*}

\begin{itemize}
    \item An open-source profiling and analysis tool that captures inter-GPU data movement with support for NCCL and NVSHMEM communication libraries
    \item Support for communication association with the involved data objects and source code origin to ease analysis and debugging
    \item User-friendly visualization tool that presents GPU communication in various levels of granularity
    \item Methods to reduce profiling overhead to 2$\times$-96$\times$ by on-device filtering, compression, and sampling  
    \item Diverse set of case studies in C/C++ and Python that covers BFS, 2D Stencil, NCCL bandwidth test, and a Deep Learning CosmoFlow \cite{cosmoflow} model in PyTorch.
 to demonstrate the different capabilities of the tool. 
\end{itemize}

While \tool{} is not intended to replace established profiling tools such as Nsight Systems, it complements them by filling a significant gap in the GPU tool portfolio. 
With a recent shift towards greater GPU autonomy, where devices directly manage communication \cite{GPU-Triggered-Networking, gio, atos}, and the widespread use of communication libraries like NCCL and NVSHMEM, \tool{} becomes an important addition to the GPU toolset. It aids both application users and developers in effectively pinpointing communication bottlenecks and optimizing their code to achieve better performance. Lastly, although \tool{}'s current implementation targets NVIDIA GPUs,
many of our contributions are architecture-agnostic and can be adapted to other GPU vendors.

\section{Motivation: Identifying Gaps}

Several profiling tools have been created for GPUs, with the main emphasis on examining the performance of computational kernels, with varying degrees of support for profiling communication calls. Table \ref{tab:related-work} lists some of these tools and their support for various aspects of communication profiling. However, we have identified certain gaps in their capabilities, including:

\begin{figure}[tp]
    \centering
    \begin{subfigure}[b]{0.45\textwidth}%
    \includegraphics[width=1\linewidth]{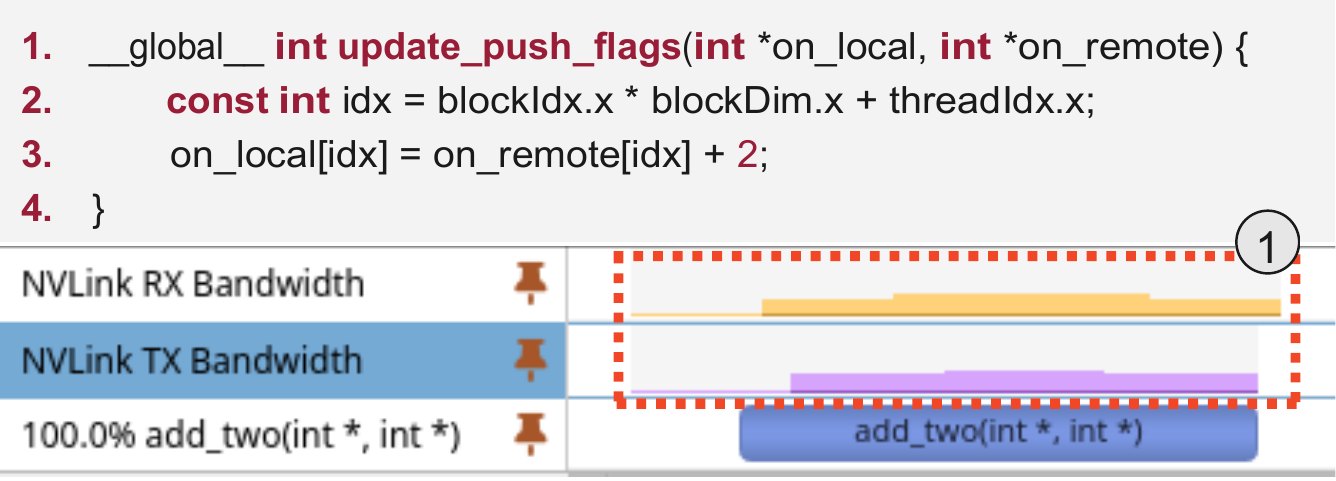}
    \caption{How P2P direct access appears in Nsight Systems}
    \label{fig:nsight-systems-p2p}

    \end{subfigure}

    \begin{subfigure}[b]{0.45\textwidth}  
    \includegraphics[width=1\linewidth]{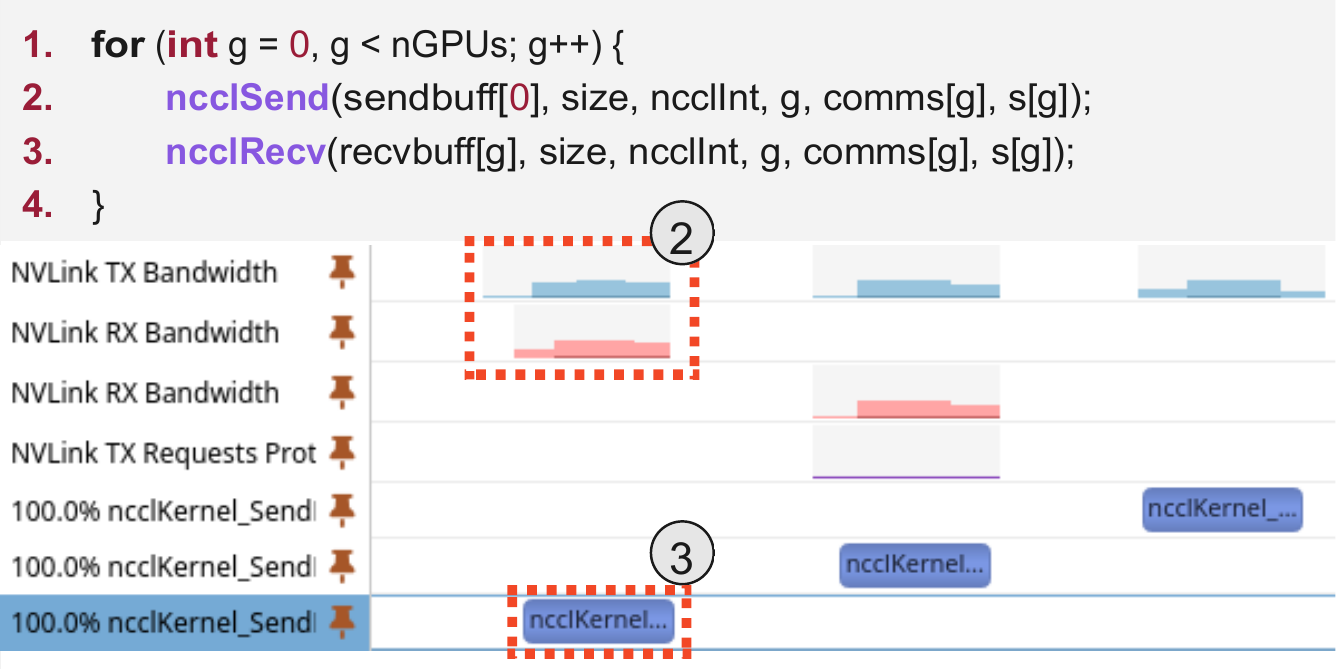}
    \caption{How NCCL Send/Receive appears in Nsight Systems}
    \label{fig:nsight-systems-device-direct-access}
    \end{subfigure}
    
    \caption{Communication profiling in Nsight System. Figures are available under CC-BY \cite{snoopie_figures}.}
    \label{fig:nsight-systems-gaps}
    \vspace{-0.5cm}
\end{figure}

\begin{itemize}
    \item Gap 1: Detecting different types of GPU-GPU communication including peer-to-peer (P2P) direct accesses to remote GPU memory 
    \item Gap 2: Tracking data movement induced by communication libraries such as NVSHMEM and NCCL 
    \item Gap 3: Ability to associate communication  with the involved data objects and  source code lines %
    \item Gap 4: Granular visualization support for data movement for analysis and debugging \\ 
\end{itemize}

{\bf Gap 1: Detecting different types of GPU-GPU communication.} 
CUDA provides several communication methods to facilitate data transfers among GPUs. The standard host-side API for data movement from one device to another is the \texttt{cudaMemcpy} family of calls, which are tracked by most existing profilers with the help of the CUDA runtime API.
The same API allows peer-to-peer communication initiated by the host through {\tt cudaMemcpyPeer} which utilizes DMA engines to perform data movement. Such communication can also be device-initiated when shared address space is available, where devices can directly load/store from/to another GPU's memory without involving the host \cite{gpu-centric}. This form of direct communication has gained prominence as more GPUs per node can be tightly connected via high-speed interconnects, facilitating low-latency access to peer GPU memories. P2P transfers also enable persistent kernels to communicate without termination, resulting in fewer kernel launches, synchronization, and reduced networking overheads \cite{Agostini_Offloading_Communication_Control_Logic, gpudirectrdma, lena-oden, gpu-centric, gravel, gio, GPU-Triggered-Networking, cpu-free}.

Despite their importance, existing profiling tools currently lack support for device-initiated transfers, and only monitor host-initiated peer transfers. For example, NVIDIA's Nsight Systems \cite{nsight-system} does not provide information on the effective addresses accessed by device-initiated remote memory operations or the devices that are the targets of the operations. Instead, its output as shown at \circled{1} in Figure \ref{fig:nsight-systems-p2p} only \emph{indicates} the occurrence of P2P accesses through a bump in NVLink traffic for a remote read. HPCToolKit \cite{hpctoolkit}, another commonly used HPC profiling tool, does not provide any information that could indicate the occurrence of P2P direct accesses.

{\bf Gap 2: Communication Library Support.}
NVIDIA provides two first-party communication libraries for multi-GPU applications: NVIDIA Collective Communications Library (NCCL) and NVSHMEM. NCCL supports both collective communication and point-to-point send/receive primitives for multi-GPU and multi-node systems. It has seen wide adoption in popular deep learning frameworks such as MxNet~\cite{mxnet}, PyTorch~\cite{pytorch}, and TensorFlow~\cite{tensorflow}, which use it to accelerate distributed training.
NVSHMEM extends the OpenSHMEM specification \cite{openshmem} to support NVIDIA GPUs and utilizes a global address space accessible with a fine-grained GPU-side API.
\cite{gpu-centric, Nvshmem_Docs}.
While Nsight Systems recognizes both NCCL and NVSHMEM calls,  it only provides timeline information. %
Figure \ref{fig:nsight-systems-device-direct-access} highlights the basic profiling information gathered by Nsight Systems at points \circled{2} and \circled{3} - showing only the timespan of NCCL calls, with no further information about the underlying properties of the operations.

{\bf Gap 3: Code Line and Object Attribution.}
Associating
profiled communication operations with their source code origin and the involved data objects can aid programmers in identifying communication-related bugs and performance bottlenecks.
Among the profiling tools listed in Table \ref{tab:related-work}, we only identified two that provide this information: Extrae \cite{extrae} and HPCToolkit.
While Extrae can attribute data movement to data objects, this feature is limited to CPUs and there is no attribution for source code lines. HPCToolkit, on the other hand, is able to attribute communication to source code lines but lacks support for objects.
Moreover, to the best of our knowledge, there are no tools that support code line and data object attributions in multi-GPU Python programs. Given the widespread use of Python in HPC and AI, such tooling becomes essential for GPU profiling in communication-heavy applications with complex codebases.

{\bf Gap 4: Data Movement Visualization.}
Current tools are generally multipurpose, and as a result, their visualization may not be specifically tailored to highlight data movement. Illustrated in Figure \ref{fig:nsight-systems-gaps}, most tools, including Nsight Systems, predominantly provide an overview of the execution time of profiled applications and offer, arguably, less informative representations of data transfers. Eztrace\cite{5948661} and ComScibe\cite{10.1007/978-3-030-71058-3_10} present a coarse-grained view of data movement in the form of a communication matrix. 
Having finer-grained information would assist users in understanding multi-GPU communication patterns, object access behaviors, and locating bottlenecks in their applications.

In summary, existing GPU profiling tools have limitations in offering a comprehensive understanding of inter-device communication in multi-GPU applications. \tool{} aims to fill the aforementioned gaps and provide an alternative perspective on communication analysis for end-users.

\begin{figure*}[th]
\begin{center}
\includegraphics[width=\textwidth]{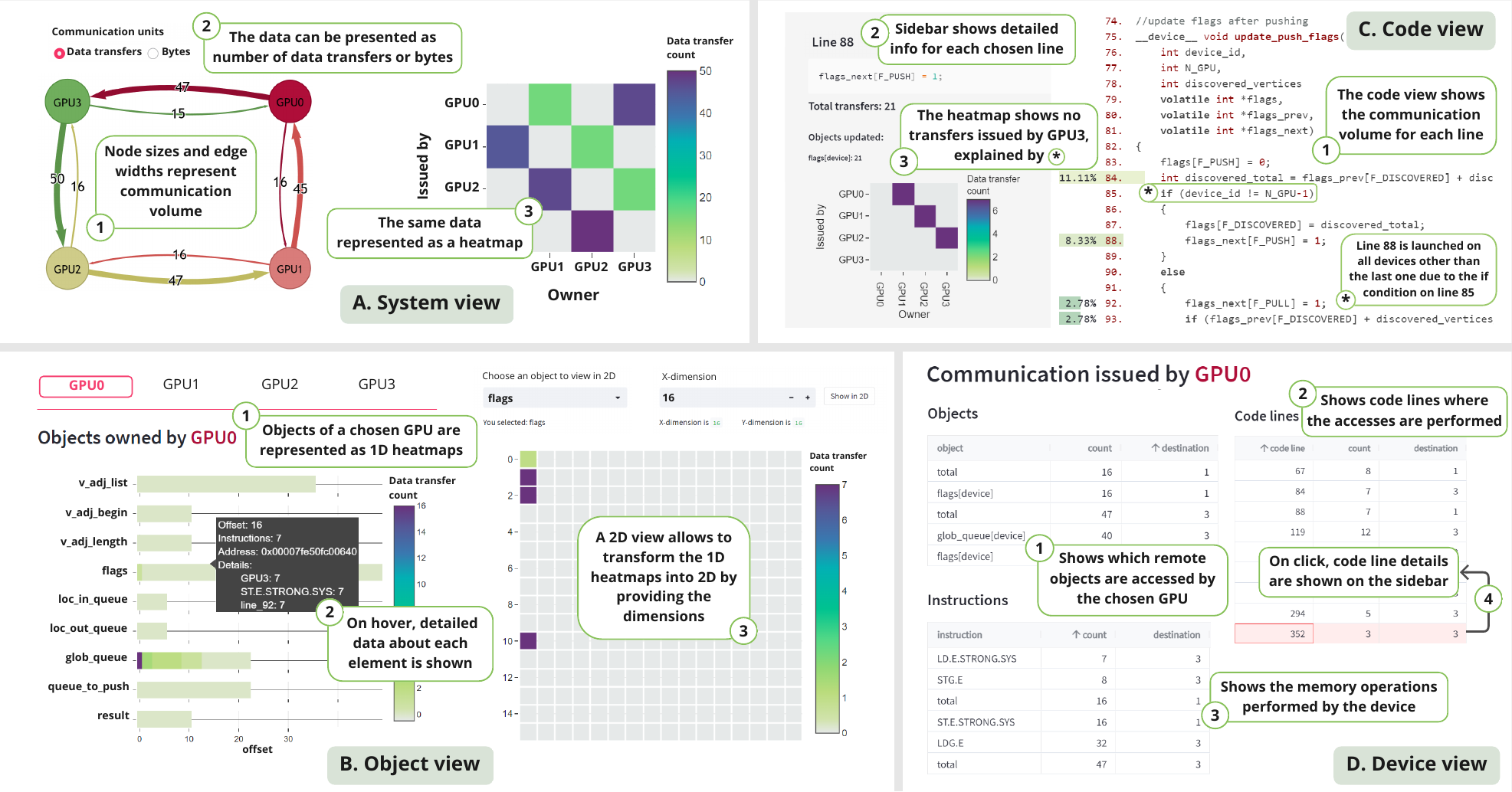}
\caption{An example of  \tool's multi-view visualizer. Figures are available under CC-BY \cite{snoopie_figures}.}
\label{fig:vis_screens}
\end{center}
\end{figure*}

\section{\tool{} with Multi-View Visualizer}
\tool{} is a binary instrumentation tool that provides a data-centric and user-friendly way to monitor and analyze device-to-device communication.
By tracking the source and destination devices of memory operations, the effective addresses, the amount of data transferred, the source code lines, and the accessed data objects, it provides a comprehensive overview of GPU communication. 

\tool{} %
provides multiple intuitively understandable ways of visually representing the data transfers. 
The visualization module presents the logged information in an interactive manner, enabling users to track and analyze the data movement between GPUs. 
It offers different levels of granularity to allow for both a coarse-grained overview of the system and a detailed view of a specific object or device.

\begin{itemize}[align=left,leftmargin=*,widest={10}]
    \item {\bf System view.} Figure \ref{fig:vis_screens}A shows the system-wide overview of communication performed by each device and between each pair of devices. The data is displayed both as an interactive graph \circled{1} and as a heatmap \circled{3}, and can be exported in textual format. Each graph node represents a device, while the edges between the nodes reflect the performed data movement. The information can be shown in terms of both the number of data transfers as well as the volume of transferred data in bytes \circled{2}.

    \item {\bf Object view.} Figure \ref{fig:vis_screens}B presents a representation of data objects on multiple granularities so that the number of operations for each specific memory address can be tracked. As can be seen in \circled{1}, each GPU has its separate object view tab. 
    On each tab, there are objects' address spaces represented as a 1D heatmap, where the information about the memory operations on each address can be viewed by hovering over the heatmap cells \circled{2}. This information includes the memory address, types of instructions performed, the remote GPU IDs participating in the communication, and source code line numbers of the memory operations. Additionally, for each object, a 2D view \circled{3} option is available. In 2D view, the user can provide the desired dimensions. 
    
    \item {\bf Code view.} Figure \ref{fig:vis_screens}C shows a view that allows to pinpoint the source code lines of the captured inter-GPU remote memory operations. It provides information about the amount of communication performed at each line \circled{1}. Code lines where communication is being performed are highlighted and the total amount of communication is shown as a percentage of the total number of data transfers. Upon clicking on a line that is involved in communication, a detailed view shown on the sidebar \circled{2} reveals the total number of remote memory operations for each line as well as a system-wide communication heatmap \circled{3} similar to the one provided in System View. Additionally, since multiple objects can be accessed in a single line, information about the objects involved in communication is provided on the sidebar as well.

    \item {\bf Device view.} Figure \ref{fig:vis_screens}D represents a brief overview of the remote memory operation types (load/store) and the accessed remote objects for each device. While the object view displays objects allocated on a selected GPU and the details regarding remote accesses of these objects by other devices, the device view complements it by instead showing the data transfers performed by the selected GPU itself. It contains object \circled{1} and instruction \circled{2} tables, representing the data access counts of remote objects and the performed instruction types. Moreover, it includes the source code line numbers \circled{3} to facilitate debugging. On click, a code line number opens the detailed view of the code line \circled{4}, the same as the one in the Code View.
    
\end{itemize}

\section{Implementation of \tool{}}

\begin{figure*}[th]
\begin{center}
\includegraphics[width=0.88\textwidth]{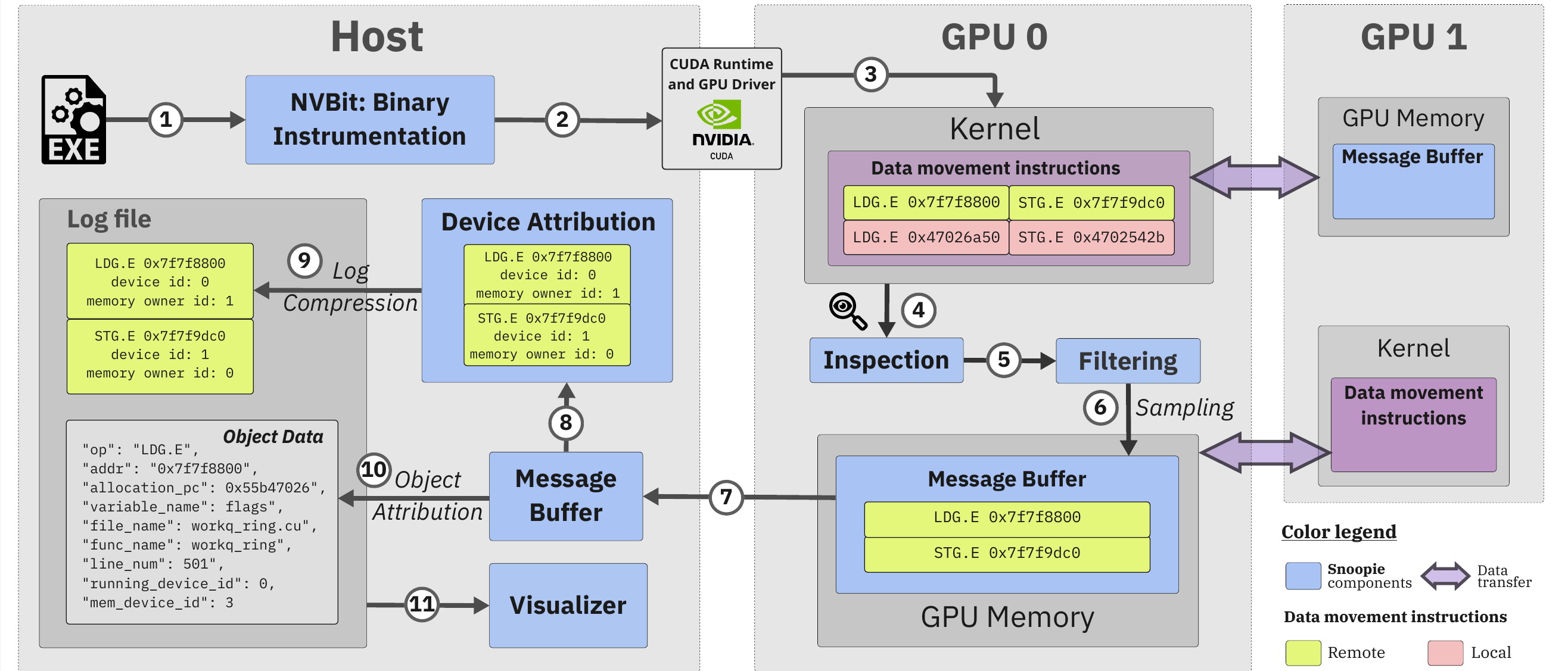}
\caption{Overview of \tool{} and its components. Figures are available under CC-BY \cite{snoopie_figures}.}
\label{fig:overview-of-snoopie}
\end{center}
\end{figure*}

Figure \ref{fig:overview-of-snoopie} shows the workflow of \tool{}. It starts by instrumenting the executable, such that subsequent CUDA runtime calls will go through \tool{} first, which is then fed to the CUDA runtime and its driver. When a kernel executes on the GPU, \tool{}'s instrumented code inspects the remote load and store instructions, which are then registered to a message buffer in the global memory of the device. 
Once the buffer is full, it is flushed to the host for processing. \tool{} attributes remote accesses to the source/destination device and associates the effective addresses with the data object in the executable. 
For other multi-GPU communication methods beyond device-initiated direct access, \tool{}  provides support for host-initiated transfers. 
To capture these transfers, \tool{} instruments the \texttt{cudaMemcpy*} family of functions. As these functions are executed on the host, no further device instrumentation is necessary. 
In the final step, \tool{} generates a log file containing trace information for offline analysis and visualization.

To develop \tool{}, one can utilize profiling interfaces like CUPTI API\cite{cupti}, Computer Sanitizer\cite{sanitizer}, or NVBit \cite{nvbit} designed for NVIDIA GPUs for data movement analysis. Key requirements for \tool{} include instrumenting direct memory accesses and tracing CUDA runtime events.
CUPTI can trace CUDA runtime events through its lightweight sampling API, but it lacks in capturing direct memory accesses. On the other hand, Compute Sanitizer, based on binary instrumentation, can monitor load/stores, including remote ones, but it operates at a module level and not at a function or kernel level. NVBit, a dynamic binary instrumentation framework, excels in granularity, allowing precise instrumentation of specific kernels or functions instead of entire modules. Its ability to trace CUDA API events and kernel launches aligns well with our requirements, making it a more suitable choice for \tool{}.

\subsection{Capturing P2P Communication}

To profile device-initiated remote memory operations, \tool{} relies on NVBit \cite{nvbit}, enabling the insertion of instrumentation code into application binary code. NVBit provides built-in functions that disassemble binary instructions, extracting details about the executed operations. Utilizing this information, we identify memory accesses, discerning between reads and writes, and capturing effective addresses. Illustrated in Figure \ref{fig:overview-of-snoopie} (\circled{1} and \circled{2}), for each identified memory access instruction, we inject a function to record the extracted information in a buffer. This injected function is invoked just before load/store operations, facilitating inspection (\circled{4}). Subsequently, local addresses are disregarded through address filtering (\circled{5}), allowing only remote addresses to proceed. In the next step, filtered addresses are sampled if the sampling feature is enabled (\circled{6}).

A channel mechanism in NVBit facilitates the communication of captured information to the host. To minimize data movement overhead, we establish the channel with double buffers in device memory. Not all threads contribute to the channel; instead, within each warp, a single thread compiles memory addresses targeted by memory operations from all other threads in the warp, creating a single message for the entire warp. Thus, each message encapsulates up to 32 memory locations, corresponding to the warp size. Upon channel saturation, the buffer is switched, and the full buffer is transmitted to host memory using a flag implementation that triggers a {\tt cudaMemcpyAsync} operation, depicted in \circled{7}.

\vspace{-0.2cm}
\subsection{Device Attribution}

The message buffer contains a list of remote addresses accessed during kernel execution, which need to be mapped to corresponding device IDs. The device attribution in \tool{} determines the device to which a pointer value belongs. This capability is facilitated by the support for Unified Virtual Addressing (UVA) provided by NVIDIA. UVA enables device pointers for different devices on the same node to share a common address space. Consequently, the {\tt cuMemAlloc} operations utilize the same address space for pointer values allocated across various devices. This shared address space ensures that if a pointer region is associated with device $A$, any subsequent allocation on device $B$ will not return an address overlapping with the region allocated on device $A$. This inherent property allows for the clear identification of the device to which a pointer belongs.
\tool{} captures memory ranges allocated for each device whenever a {\tt cuMemAlloc} call is issued by the executable to the CUDA driver and the device ID of the context in which the call is made.

To perform device attribution, \tool{}  parses each message received via the message buffer, as can be seen in Figure \ref{fig:overview-of-snoopie} in \circled{8}, and determines the region to which each pointer belongs by comparing it with the data obtained from the instrumented {\tt cuMemAlloc} operations.
Load/store operations that are a null or empty pointer value are ignored. This is because some threads within a given warp might not be performing a memory operation. The filtered information is then converted into CSV format and is fed into a Zstandard compression stream 
before it is pushed to the log to be used by the visualizer.

\tool{} uses the approach described above instead of relying on {\tt cuPointerGetAttribute} to obtain information about pointers received from the device. While this CUDA API can provide useful information about a pointer, including its device, base address, and memory type, it may not be able to determine the device ordinal 
if the memory pointed to by the pointer has already been freed.
Our alternative approach enables \tool{} to overcome this limitation by leveraging memory range to device mappings, which persist beyond the lifetime of the memory allocation itself.
However, for host-initiated transfers, \tool{} employs {\tt cuPointerGetAttribute} for device attribution because \tool{} can guarantee that the instrumentation of {\tt cudaMemcpy*} functions occurs before any pointer is deallocated, allowing for accurate device attribution in these scenarios.

\subsection{NCCL and NVSHMEM Support}

\begin{figure}[t]
    \centering
    \includegraphics[width=\linewidth]{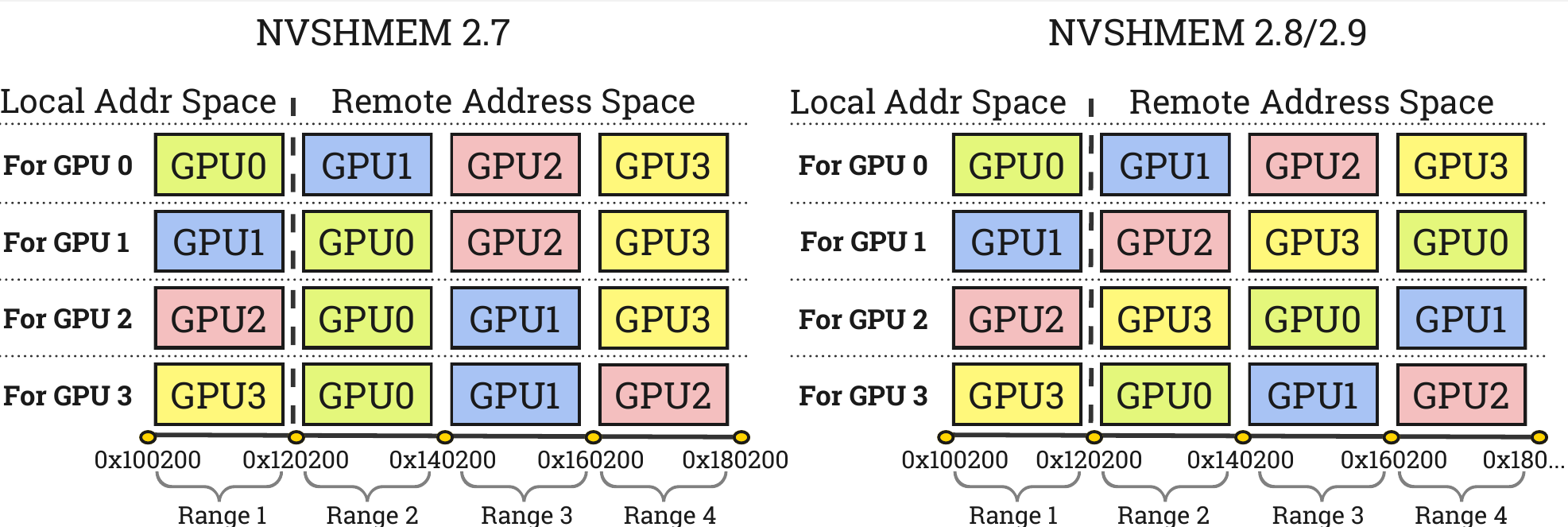}
    \caption{NVSHMEM address space. Figures are available under CC-BY \cite{snoopie_figures}.}
    \label{fig:nvshmem-mem-spacing}
\end{figure}

As NCCL and NVSHMEM are becoming more popular for scaling deep learning and HPC applications to multiple GPUs, profiling their communication is essential. 
NCCL performs communication using device-initiated direct accesses, as a result, \tool{} can support NCCL directly without requiring special handling. 
Even though NVSHMEM also utilizes device-initiated direct accesses for many of its communication calls, 
it demands special handling due to its symmetric memory layout and the way peers communicate with each other. 

In NVSHMEM, each device has a base address for its address space, and it designates a portion of this address space for each peer. However, the designations for each peer's address space may not match between peers. Figure \ref{fig:nvshmem-mem-spacing} illustrates the virtual address spaces of each GPU to access other peers within the same node for NVSHMEM v2.7, 2.8 and 2.9. For example, in NVSHMEM 2.7, accessing GPU2 from GPU1 would appear as a load/store operation in Range 3, while accessing the same variable on GPU2 from GPU3 would appear as a load/store operation in Range 4. This method of segmenting the address space makes it non-trivial for our proposed approach to work, as multiple addresses point to the same physical address.

To resolve this issue and add device attribution support for NVSHMEM, \tool{} captures the identifier of the NVSHMEM Processing Element (PE) responsible for executing the direct access operation. An NVSHMEM PE refers to the process that forms an NVSHMEM job. By utilizing the PE identifier along with the captured memory address, \tool{} is able to determine the device on which the respective memory address resides
by taking the address spacing rules illustrated in Figure \ref{fig:nvshmem-mem-spacing} into account.

\subsection{Source Code Line Attribution}
\label{sec:src-code-line}

To associate inter-GPU memory operations to their respective locations in kernel source code, \tool{} employs NVbit's {\tt nvbit\_get\_line\_info} function to extract line information from each instrumented binary instruction. However, it should be noted that NVBit faces limitations in extracting this information from a significant portion of instructions.
To overcome this limitation, \tool{} additionally utilizes \texttt{nvdisasm} \cite{cubin} to obtain mapping information between PTX assembly instructions and CUDA source code lines. Given that NVBit supports the disassembly of instrumented instruction binaries into their SASS assembly instructions, \tool{} leverages this mapping information by matching a disassembled SASS instruction with its PTX version, which has been previously mapped to a source code line in the output of \texttt{nvdisasm}. This pattern matching between SASS and PTX instructions is only performed when {\tt nvbit\_get\_line\_info} fails to extract the code line information.

\subsection{Data Object Attribution}

To identify data objects that are accessed during communication, we developed \tool{} with the data object attribution ability. This feature consists of two phases: (1)  object recording phase, and (2) object mapping phase. 

\subsubsection{Object recording phase} 
\begin{figure}[tp]
\centering
\includegraphics[width=0.9\linewidth]{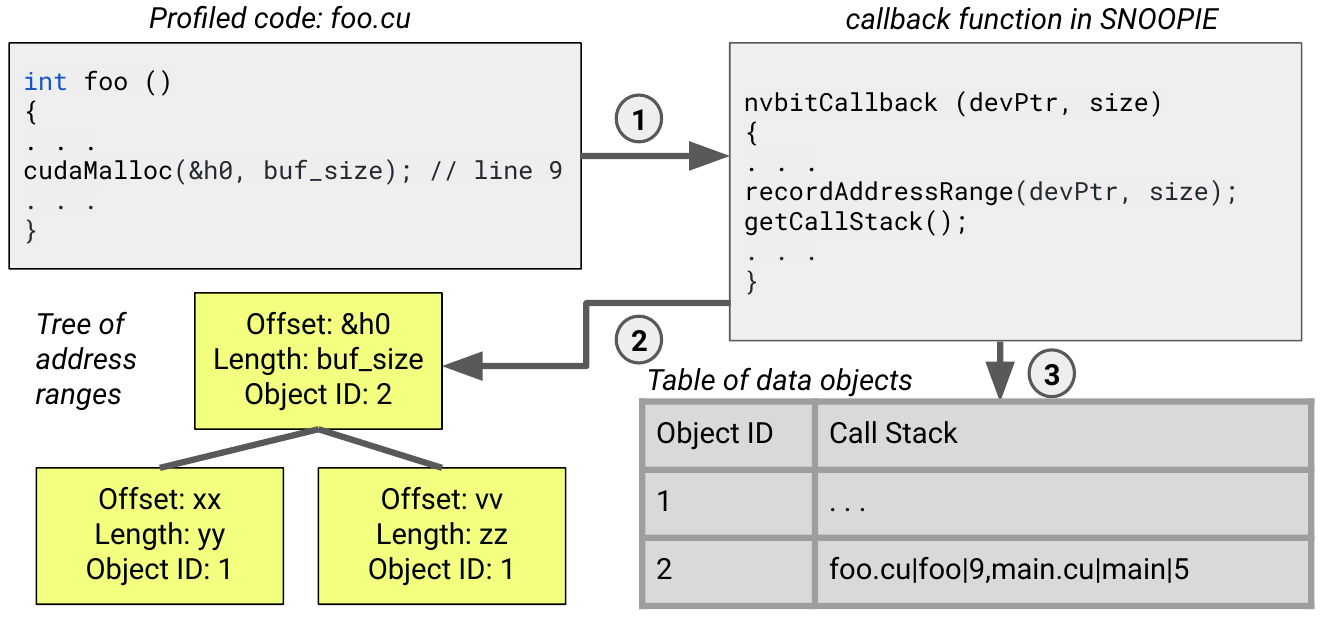}
\caption{Object recording phase of data object attribution. Figures are available under CC-BY \cite{snoopie_figures}.}
\label{fig:object-attribution-workflow}
\end{figure}

The object recording phase is illustrated in Figure~\ref{fig:object-attribution-workflow}.
During the application execution, \tool{} intercepts function calls for dynamic memory allocations such as \texttt{cudaMalloc} and \texttt{cudaMallocHost} using a callback function that is inserted to the address space of the profiled process using the \texttt{LD\_PRELOAD} utility\cite{LDPRELOAD-www} in Linux. 
This callback function calls the original allocation function and records the offset address and byte length of the memory region in a global splay tree (\textit{tree of address ranges}), where each node corresponds to an address range generated by the memory allocation call as shown in Figure~\ref{fig:object-attribution-workflow} \circled{2}. In addition to these pair of offset and byte lengths, each node also records an object ID that uniquely identifies the call stack information of the memory allocation call. 

We define a data object in our profiling as the call stack of a memory allocation call, encompassing information about the file name, function name, and line number. According to our definition, a data object can be associated with multiple non-contiguous memory regions. This situation arises when there are multiple memory allocation calls sharing the same call stacks, such as a {\tt cudaMalloc} being repetitively called within a loop.
To mitigate redundancy in the \textit{tree of address ranges}, we store the details of memory allocation call stacks in a separate hash table, the \textit{table of data objects}, where each entry corresponds to a unique call stack, identified by an object ID that serves as the primary key.

To enhance the debugging process, we have introduced custom wrapper allocation functions enabling programmers to include the variable name, associated with the address range, as a function argument. With this additional information, \tool{} incorporates variable names into the visualization, aiding users in a clearer understanding of object attribution.

\subsubsection{Object mapping phase} \tool{} intercepts memory operations during the runtime of profiled applications, logging the memory addresses accessed in these operations. Upon the completion of an application run, \tool{} processes the logged data. At this stage, each recorded memory address undergoes a query in the tree of address ranges to determine the corresponding address range it belongs to. Once \tool{} identifies the relevant node in the  tree, it utilizes the object ID within the node to query the corresponding object in the \textit{table of data objects}. Subsequently, \tool{} includes the retrieved entry in the log file.

\subsection{Support for Profiling Python Programs}
\tool{} supports communication detection for Python applications, with specific considerations for code line and object attributions. These attributions are presently available for Python codes using Numba~\cite{numba} or PyTorch~\cite{numba}.
For code line attribution in Numba-decorated functions, \tool{} leverages Numba's feature that generates line info in dispatched CUDA kernels. By enabling this feature, \tool{} can extract line info for each instruction using NvBit's 
 \texttt{nvbit\_get \_line\_info}.
To extend support for source code line attribution in PyTorch programs, \tool{} captures the call stacks of profiled Python code during the interception of NCCL kernel launches within PyTorch. This information enables \tool{} to attribute the communication detected in the launched NCCL kernels to the specific lines in the Python code that trigger the kernel launches.

For data object attribution, \tool{} logs the allocated address ranges by instrumenting calls to the memory allocation functions of Numba and PyTorch. To enable the instrumentation of Python function calls within \tool{}, we employ the pybind11 library and the C API of Python. This combination facilitates interoperability between Python libraries and the C++ code of \tool{}.
\\

\subsection{Reducing Profiling Overhead}
Binary instrumentation and monitoring every memory access not surprisingly leads to runtime overhead. We implement several methods to lower this overhead. 

\begin{itemize}[align=left,leftmargin=*,widest={10}]

\item {\bf On-Device Filtering}: 
Initially, \tool{} transmitted all memory operation records from the instrumented functions to the host. Subsequently, the host filtered these records to retain only the remote accesses. However, this approach posed challenges for certain applications, like stencil, where non-remote memory operations significantly outnumbered remote ones.  To overcome this issue, \tool{} performs the filtering directly on the device-side and sends only the remote records to the host, eliminating unnecessary data movement. The filtering is done by the first active thread within the warp, responsible for placing addresses into the communication channel with the host.
Instead of placing all addresses onto the channel, the warp's first active thread checks each address against an on-device maintained list of memory allocations that occurred on the host and the devices they belong to. If a remote address is found, the group of addresses are placed on the channel. Otherwise, the addresses are discarded. 

\item {\bf Sampling}:
To minimize runtime overhead and decrease log file size further, we incorporated an optional sampling technique for capturing remote memory operations. This method is integrated into the callback function, which is triggered when a remote memory operation takes place on a GPU. Within the callback function, a pseudo-random number is generated and compared to a threshold specified by the user. If the generated number is lower than the threshold, the callback function records the intercepted memory operation and sends it to the message buffer, as shown in \circled{6} in Figure \ref{fig:overview-of-snoopie}. Conversely, if the generated number exceeds the threshold, the intercepted operation is discarded. 
Note that sampling is performed after on-device filtering so as not to adversely affect the accuracy. 

\item {\bf Log Compression}: 
As the tool can monitor and record millions of memory operations, we utilize compression to reduce the size of the logs. This results in a significant reduction, shrinking the accumulating logs from 0.5 gigabytes for 5 million records down to 7 megabytes. 

\end{itemize}

Moreover, the programmer can reduce the overhead by narrowing down the scope of profiling either by selecting a specific device function to monitor or by isolating the communication code to its own device function. 
A user can also wrap {\tt cudaMalloc} calls for data objects of interest so that \tool{} only monitors those memory regions. 

\subsection{Discussions on Limitations}

\tool{} currently focuses on intra-node multi-GPU setups, meaning it does not support NVSHMEM usage in a multi-node GPU cluster at this time. A similar constraint applies to NCCL applications that utilize MPI; \tool{} only supports NCCL when used within a single OS process. This limitation also impacts future compatibility with PyTorch distributed training applications that rely on NCCL with MPI, underscoring the importance of supporting MPI-utilizing applications in our upcoming efforts. However, this limitation does not apply to NVSHMEM for multi-process single-node usage. Nonetheless, we anticipate no technical obstacles in extending \tool{} to multi-node configurations in the future.

Timeline information falls outside the scope of \tool{}. Its intended use complements the existing GPU tooling landscape, which already includes timeline information through tools like Nsight Systems.

While \tool{} is developed for the CUDA environment at this time, it can be extended to support AMD GPUs.
ROCm provides analogous methods of GPU-initiated communication with both P2P RDMA \cite{rocm-docs} and GPU-centric APIs - RCCL (NCCL equivalent) \cite{rccl-docs} and ROC\_SHMEM \cite{roc-shmem} (NVSHMEM equivalent),
and supports runtime and binary instrumentation \cite{omnitrace}.

\section{Evaluation}

\begin{figure}[t]
    \begin{center}
    \begin{subfigure}[t]{0.45\textwidth}
        \includegraphics[width=\textwidth]{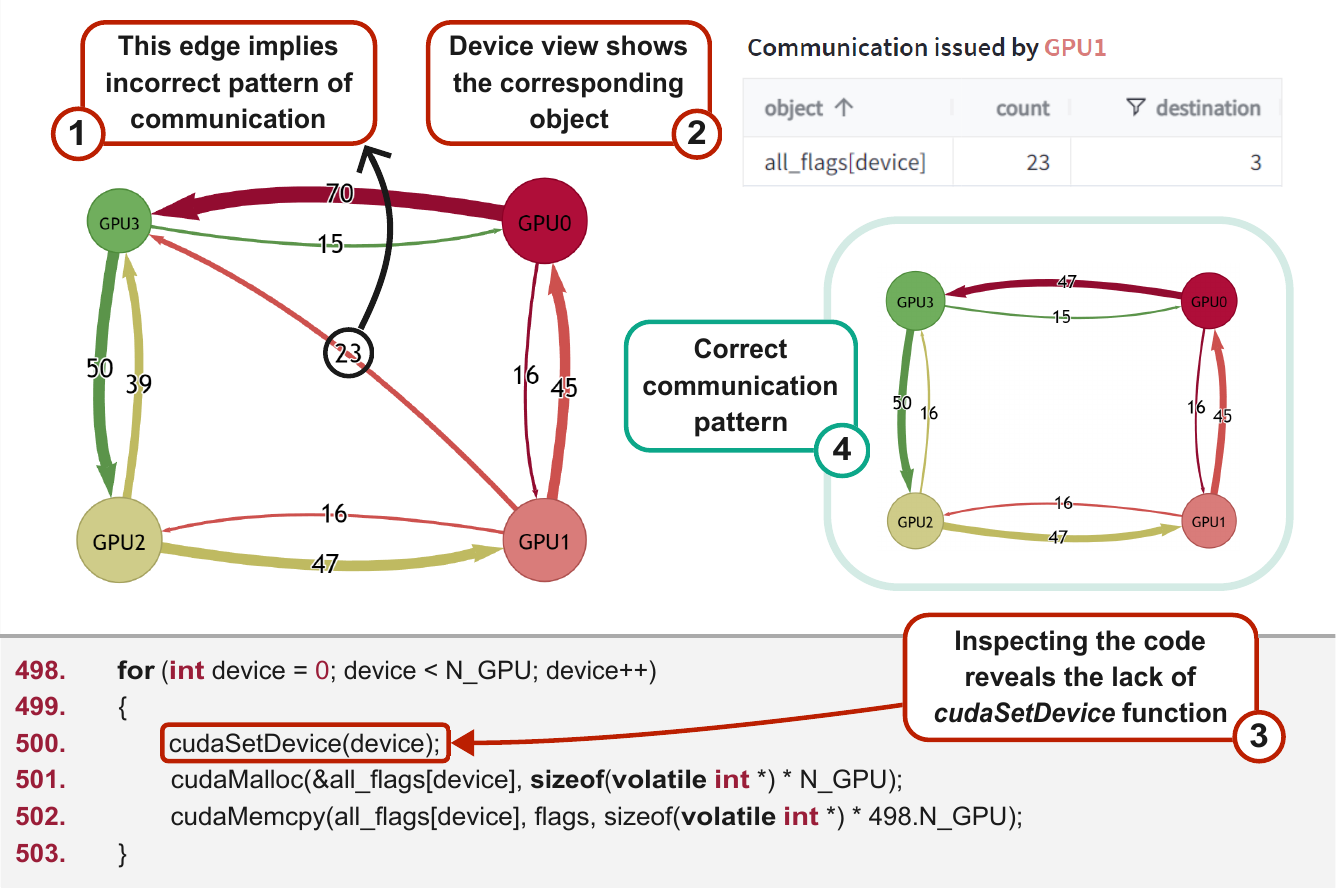}

        \caption{Demo 1: Fixing a bug with the help of \tool{}}
        \label{fig:demo1}
    \end{subfigure}
    \hspace{0.05\textwidth}
    \begin{subfigure}[t]{0.45\textwidth}  
        \includegraphics[width=\textwidth]{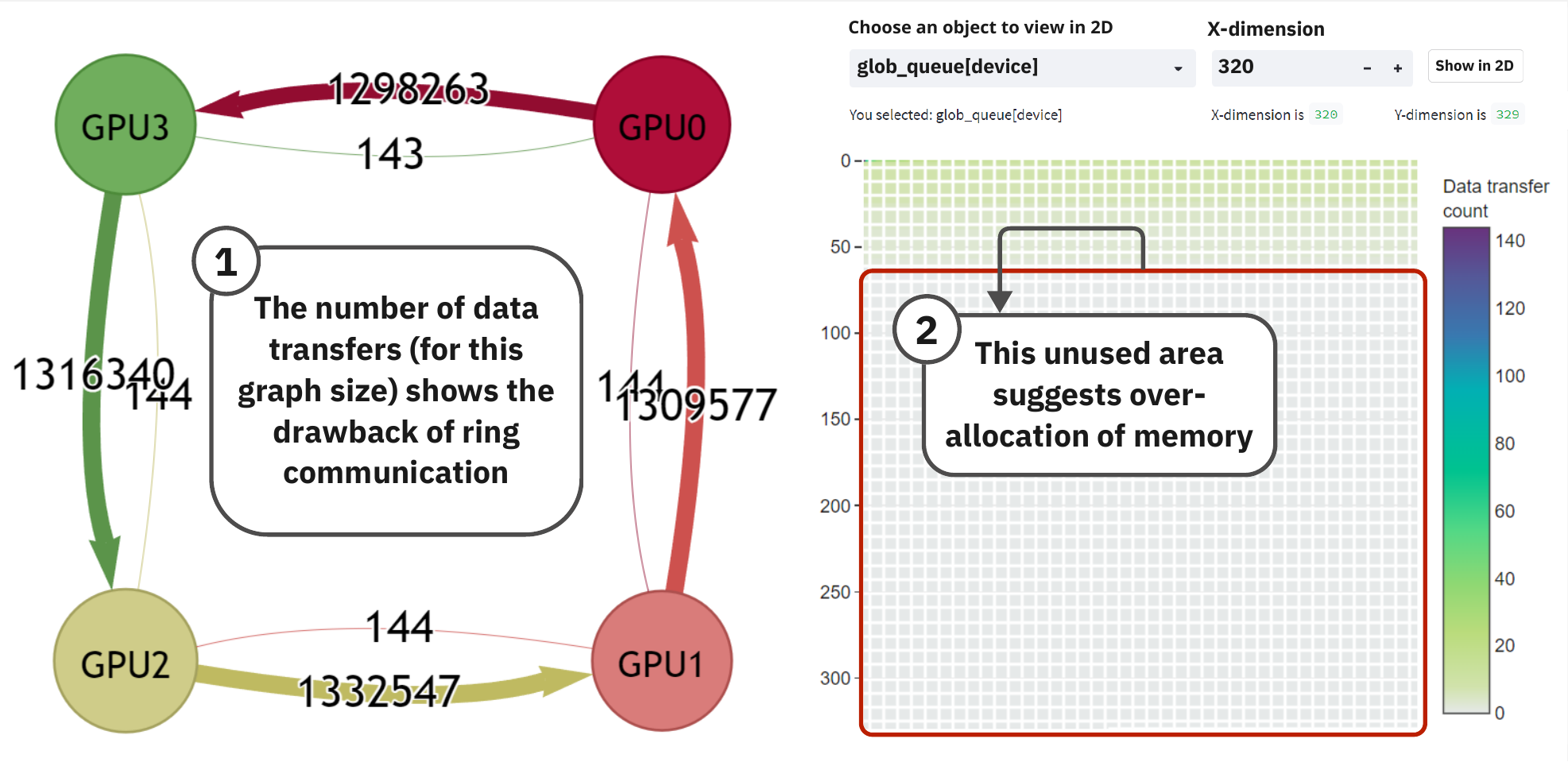}

        \caption{Demo 2: Analysis of a ring execution model}
        \label{fig:demo2}
    \end{subfigure}
    \caption{Use cases of \tool{}: a BFS example. Figures are available under CC-BY \cite{snoopie_figures}.}
    \label{fig:bfs}
    \end{center}
\end{figure}

This section evaluates the effectiveness of \tool{} and demonstrates its usability, accuracy, debugging aid, and overhead with four use-cases. The evaluation will focus on Breadth-First Search (BFS), 2D Stencil Jacobi, an AllReduce benchmark from NCCL tests\cite{nccl-tests}, and CosmoFlow \cite{cosmoflow}
The experiments are conducted on a node with 8x Nvidia Ampere 100 GPUs, with an AMD 7763 64-core processor.
All the experiments are done without sampling enabled
and with device-side filtering unless otherwise stated. 

\subsection{Breath-First Search (BFS) }
BFS traversal is an algorithm that is commonly used to traverse and search graph-like data structures. Despite following a simple principle of traversing nodes level by level, %
it has irregular communication patterns.
Therefore, BFS is particularly useful for analyzing communication. 
 The BFS implementation in use is based on a ring principle and leverages the device-initiated P2P direct remote memory access to communicate between the GPUs. Devices exchange data in a ring manner by reading from globally accessible buffers, while the order is ensured by passing flags. 
 
{\bf Demo 1. Observing the communication patterns and debugging.}
This demonstration is run on a small graph containing only 12 nodes for better traceability. The communication pattern in a system-wide view on Figure \ref{fig:demo1} shows that mostly the neighboring devices are communicating with each other. Yet, there is an arrow \circled{1} representing communication between devices 1 and 3, which are not neighbors, thus not expected in a ring communication. The object view and the device view can give further hints on the possible reason for such behavior \circled{2}, highlighting an object associated with the excess communication and pointing out that the data is being \emph{read} from GPU3. Closer inspection \circled{3} of the object's allocation reveals that although all the objects are allocated on their corresponding devices correctly, the pointer array to all devices' flags is allocated only on GPU3. Therefore, all the flag updates incorrectly require getting the flag pointer from GPU3 by all the devices in the system. Proper allocation of the pointer array solves the issue \circled{4} and removes the excess transfers.  

{\bf Demo 2. Performance bottleneck detection.} For a use case that involves analysis of issues on a larger scale we use a bigger graph \emph{web-BerkStan} with around 670K nodes. %
Figure \ref{fig:demo2} shows the \tool{} output. The resulting amount of communication in the system turns out to be almost twice the number of vertices. The sheer amount of transferred data is not problematic, since communicating the node data requires transferring 2 values: the node index and its depth. However, all communications need to pass through a ring \circled{1} yielding excess transfers. This fact suggests usage of direct transfers between the devices with buffers for each device pair at the cost of finding the nodes' owners. Moreover, inspection of the ring buffer in \circled{2} shows that the allocated buffer space for this graph is much larger than required, suggesting memory consumption for communication buffers could be reduced.

\begin{figure}[tp]
    \centering
    \includegraphics[width=1\linewidth]{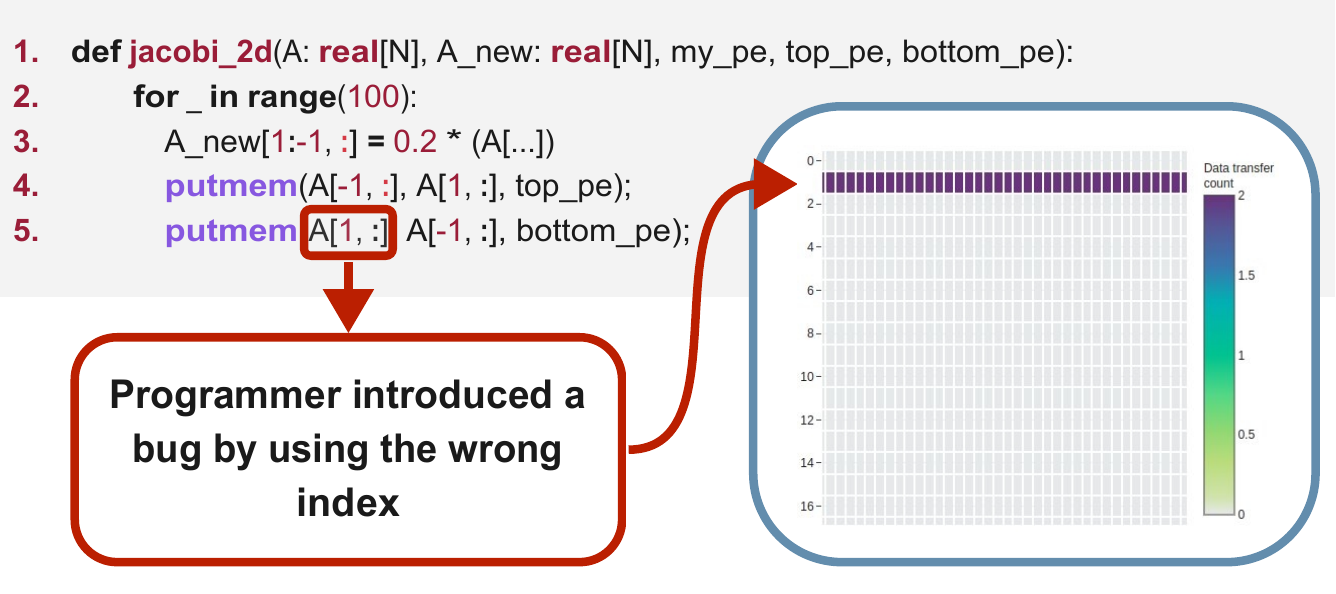}
    \caption{Demo 3: Debugging an erroneous halo region of a 2D  stencil computation using \tool{}'s object view. Instead of C++, simplified Python code is shown for clarity. Figures are available under CC-BY \cite{snoopie_figures}.}
    \label{fig:demo3}
\end{figure}

\subsection{2D Stencil Computation}

 The stencil computation involves updating the value of each element on a grid based on the values of its neighboring elements, typically within a fixed-size window.
 These computations are commonly found in scientific applications and image processing. 
Our evaluation uses a 5-point 2D star stencil operation over a domain of $1024^2$ which is divided over 8 GPUs on the y-axis such that each GPU operates on an array of size $1024*128$.  We experimented with two variants of the kernel; i) {\em Stencil-p2p} is based on the implementation by NVIDIA \cite{nvidia-multi-gpu-programming-models-p2p} that uses the device-initiated P2P communication model across GPUs. 
ii) {\em Stencil-NVSHMEM} is same as Stencil-p2p, however, uses NVSHMEM implementation provided by NVIDIA \cite{nvidia-multi-gpu-programming-models-nvshmem}.

{\bf Demo 3. Using object view to debug communication.} 
To simulate a typical bug in stencil computations, we deliberately introduce an indexing error in the Stencil-p2p code. This error involves changing the halo region in each GPU, which corresponds to the region accessed by the neighboring GPU. The object view  of \tool{} demonstrates the data transfers visually and shows all the data objects allocated in each GPU. Notably, there are two one-dimensional arrays, named \texttt{A} and \texttt{A\_new}, which are accessed as 2D arrays in the stencil computation. The intended halo regions for these arrays in each GPU are the first and last rows. However, as depicted in Figure \ref{fig:demo3}, the color-coded remotely accessed region should be the first row of the \texttt{A\_new} array, but mistakenly, it is indicating the second row.

\begin{figure}[tp]
    \begin{subfigure}[b]{0.23\textwidth}
    \includegraphics[width=1.0\linewidth]{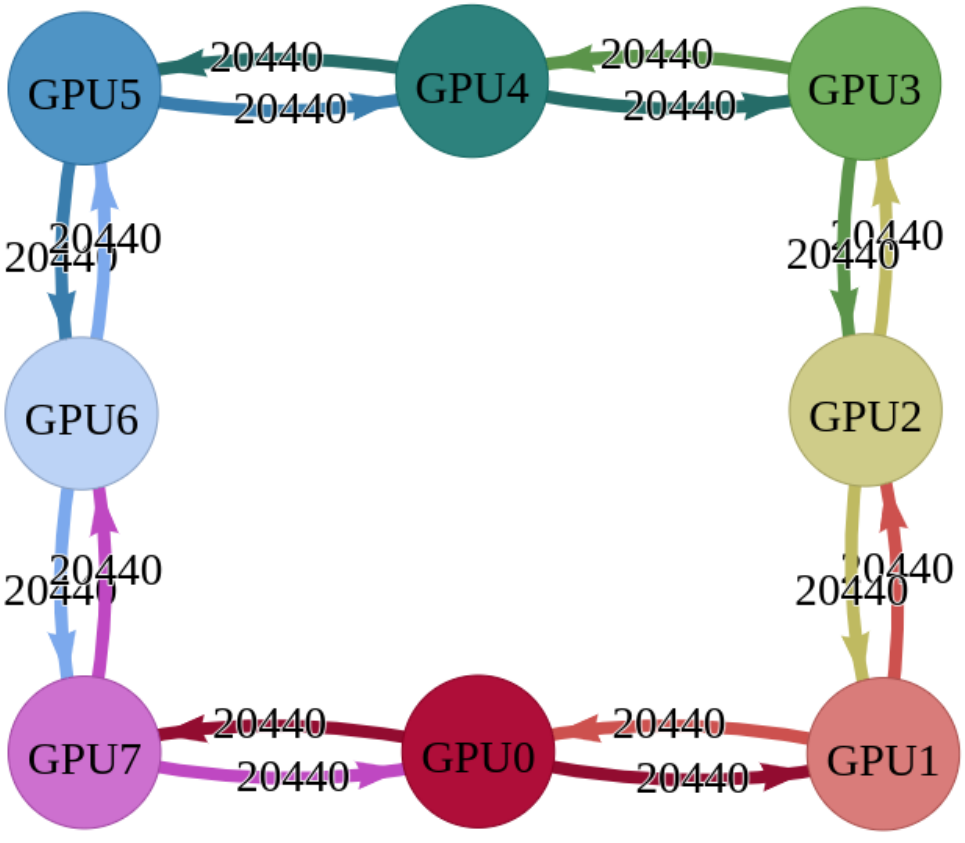}
    \caption{Stencil-p2p}
    \label{fig:demo4-stencil-p2p}
    \end{subfigure}
    \begin{subfigure}[b]{0.23\textwidth} 
    \includegraphics[width=0.95\linewidth]{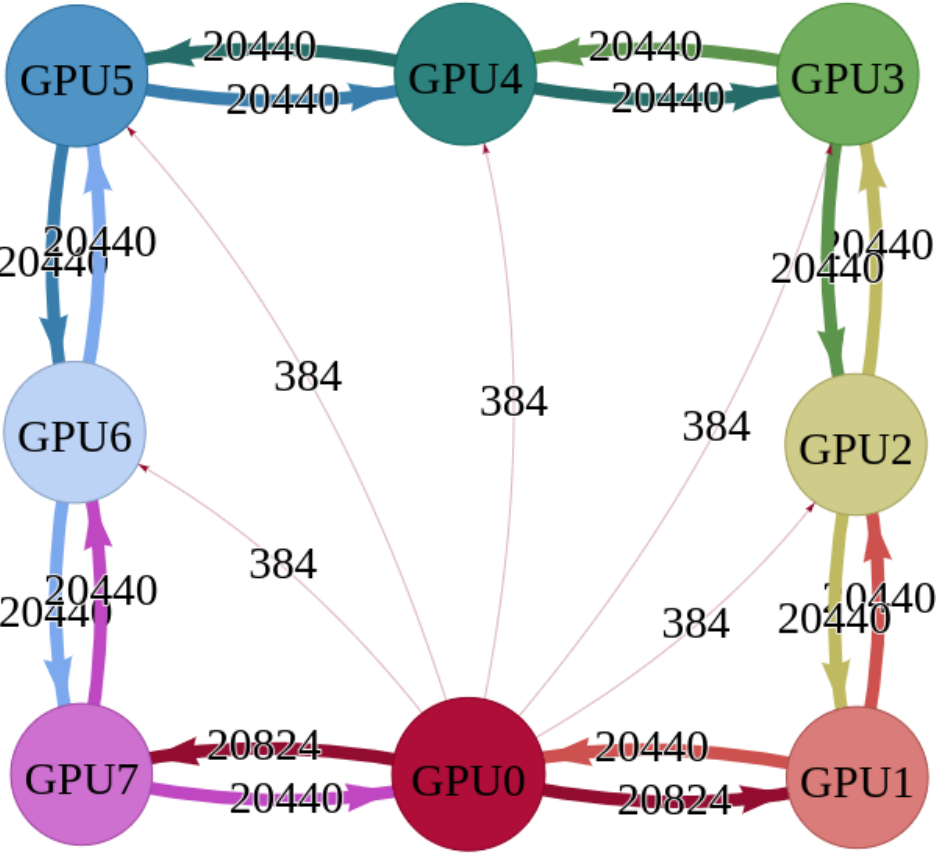}
    \caption{Stencil-nvshmem}
    \label{fig:demo4-nvshmem-2.9}
    \end{subfigure}
\caption{Demo 4: Comparing communication pattern and volume across two multi-GPU stencil implementations. Figures are available under CC-BY \cite{snoopie_figures}.}
    \label{fig:demo4}
\end{figure}

{\bf Demo 4. Comparing communication induced by different stencil implementations.}
Figures~\ref{fig:demo4-stencil-p2p} and \ref{fig:demo4-nvshmem-2.9} compare the communication patterns and volumes of the Stencil-p2p and Stencil-NVSHMEM codes on 8 GPUs. 
Both implementations display nearly the same communication patterns and volumes. The only difference is the small-sized directed edges from GPU 0 to the rest of the GPUs in the NVSHMEM version. This small-volume data traffic is caused by the host-initiated inter-GPU communications using device-to-device \texttt{cuMemcpy} function calls.

\subsection{AllReduce from NCCL}

\begin{figure}[t]
    \centering
    \includegraphics[width=\linewidth]{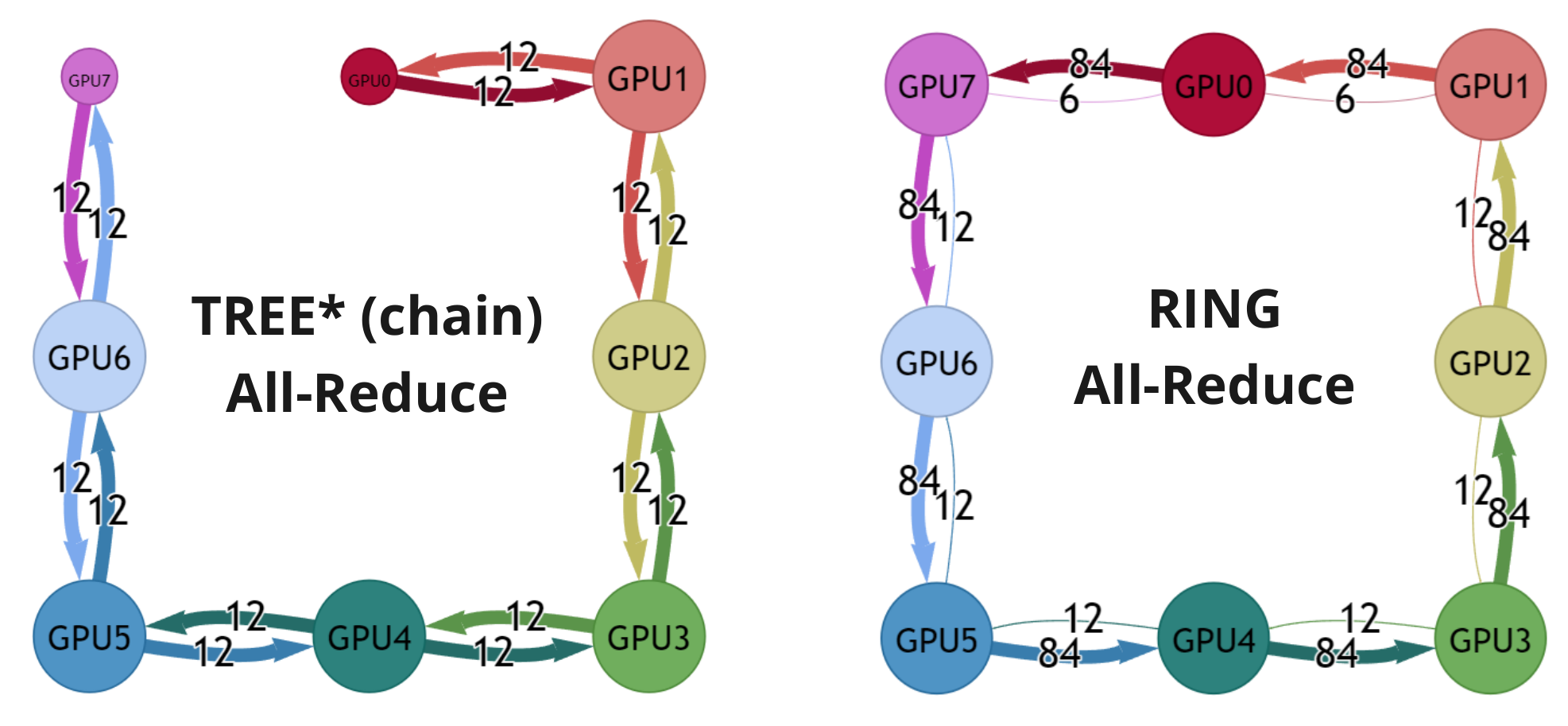}
    \caption{Demo 5: Comparing NCCL All-Reduce with Tree or Ring algorithms. Figures are available under CC-BY \cite{snoopie_figures}.}
    \label{fig:nccl-allreduce}
\end{figure}

NCCL is a communication library that uses different underlying algorithms depending on the supplied load. In this demo, we focus on one collective communication primitive, namely \texttt{AllReduce}, which offers different algorithms including {\tt RING} and {\tt TREE}. The {\tt TREE} algorithm, introduced in NCCL 2.4, is designed to improve scalability and performance for small and medium-sized operations by using a double binary tree which appears as a chain/list when running in a single node to implement {\tt all\_reduce}, with each rank sending and receiving $2N$, except for the edges which exchange $N$ where $N$ represents the size of the data being reduced. The {\tt RING} algorithm, on the other hand, transmits data in chunks around the ring, with each rank sending data to the next and receiving data from the previous rank to achieve maximum bandwidth. As NCCL can switch between algorithms based on latency and bandwidth requirements of the target platform, it might be valuable for the user to know which algorithm is being used for the communication operation.

{\bf Demo 5. NCCL communication visualization.} 
We ran the {\tt all\_reduce} NCCL test code with the output shown in Figure \ref{fig:nccl-allreduce}. 
The system view of \tool{} enables users to observe 
the chain structure when the {\tt TREE} algorithm is used. Note that this is the expected behavior of {\tt TREE}  within a single node. The visualizer also displays the communication size of each GPU node, indicating how GPU0 and GPU7 are smaller than the other GPUs when using the {\tt TREE} algorithm, as expected for the ends of the chain. In contrast, the visualizer shows almost equivalent sizes for all nodes when using the {\tt RING} algorithm. 

\subsection{CosmoFlow}

CosmoFlow is a distributed deep learning application built to predict cosmological parameters
given 3D matter distribution data \cite{cosmoflow}. In order to ensure compatibility with \tool{} and to show
different possible communication patterns, we port the PyTorch model to use a single process and implement two versions -
Data Parallel and Model Parallel.

\begin{figure}[tp]
    \begin{subfigure}[b]{0.23\textwidth}
    \includegraphics[width=1.0\linewidth]{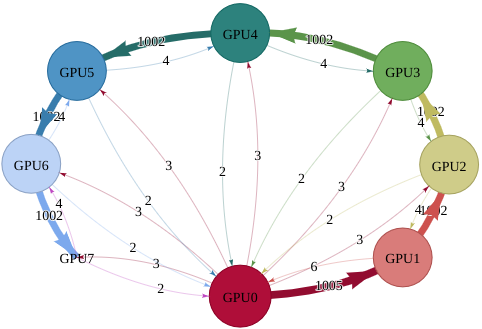}
    \caption{Data Parallel}
    \label{fig:demo7-cosmoflow}
    \end{subfigure}
    \begin{subfigure}[b]{0.23\textwidth} 
    \includegraphics[width=1.0\linewidth]{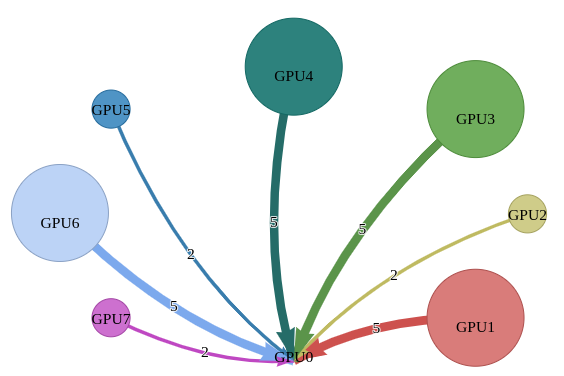}
    \caption{Model Parallel}
    \label{fig:demo7-cosmoflow-matrix}
    \end{subfigure}
\caption{Demo 6: PyTorch implementations of Cosmoflow utilizing different parallelization strategies. Figures are available under CC-BY \cite{snoopie_figures}.}
    \label{fig:demo7}
\end{figure}

{\bf Demo 6: Comparing data vs model parallelism.} The Data Parallel implementation utilizes built-in automatic parallelism functionality in PyTorch and splits the model and the data across devices.
The communication generated uses NCCL and we observe a balanced chain shown in Figure \ref{fig:demo7-cosmoflow}.

Our naive Model Parallel implementation, on the other hand, requires manual assignment of model layers to devices and is more susceptible to issues with balancing. Our port uses a naive strategy of sequentially assigning layers to subsequent devices. 
In contrast to the Data Parallel implementation that splits data across devices, the Model Parallel implementation requires the placement of all tensors in a single GPU, which in our case is GPU 0. The centralized location of allocated memory in a single GPU causes the model layers in all of the other devices to read/write into the memory of GPU 0 as shown by the pattern in Figure \ref{fig:demo7-cosmoflow-matrix}.
Through this experiment, we note that \tool{} can greatly aid in mapping communication among devices and comparing the communication patterns between different parallelization strategies.

\subsection{Overhead and Accuracy}

\begin{figure}[t]
    \centering
    \includegraphics[width=\linewidth]{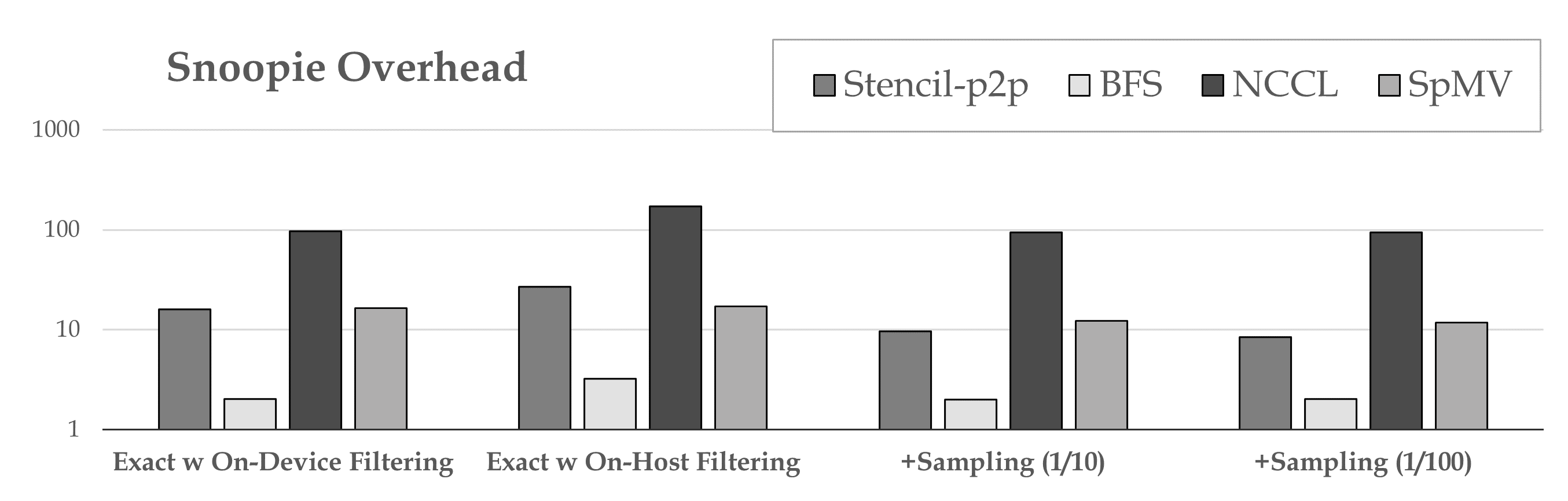}
    \caption{Overhead comparison of \tool{} compared to uninstrumented code. The Y-axis is log-scaled. Figures are available under CC-BY \cite{snoopie_figures}.}
    \label{fig:overhead}
\end{figure}

\tool{} is accurate, but being an instrumentation-based tool, it introduces runtime overhead to the profiled application. Figure \ref{fig:overhead} shows the performance slowdown compared to the baseline without instrumentation. Note that Overhead measurements include the log compression time.
The overhead of \tool{} is much higher on NCCL due to the huge number of memory allocations on GPUs that NCCL does in its implementation of collective communication functions. Since \tool{} captures the allocated address ranges to detect inter-GPU remote accesses, this high number of memory allocations affects the profiling overhead.
Performing device-side filtering of the remote operation results in a 1.8$\times$ improvement compared to filtering on the host, without compromising the profiling accuracy. This optimization also applies to other use cases, albeit to a lesser extent. Specifically, when the device-side filtering optimization is applied, the overhead of \tool{} is $16\times$, $2.04\times$, $96\times$, and $16.4\times$ for Stencil, BFS, and NCCL, respectively.

To further reduce overhead, sampling-based profiling can be employed but may lower \tool{}'s accuracy. This approach shows much more impact on stencil than the other benchmarks due to \tool{}'s memory-intensive nature and numerous local loads/stores, which \tool{} discards when these accesses are not sampled. While the overhead of on-device filtering, 1/10 sampling rate, and 1/100 sampling rate are nearly the same for BFS, NCCL, and SpMV, the overhead on stencil shows gradual reduction with an overhead of $9.6\times$ on 1/10 sampling rate and $8.5\times$ on 1/100 sampling rate.

There are cases in which the majority of the overhead was caused by the massive size of communication reported and the processing and logging following it. In these cases, adjusting the sampling parameter and the profiling method did not achieve notable improvements, as the overhead ranged between $10\times$ and $20\times$ though we argue that the real execution time is reasonable.

\begin{figure}[tp]
    \begin{subfigure}[b]{0.23\textwidth}
    \includegraphics[width=0.9\linewidth]{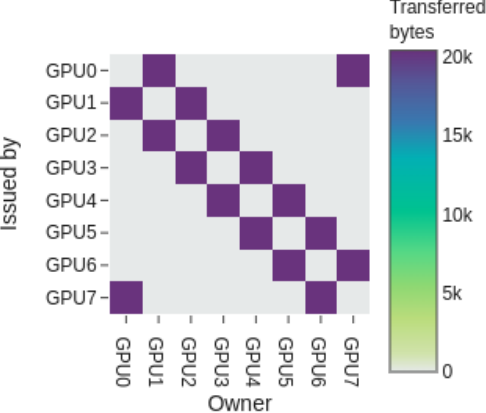}
    \end{subfigure}
    \begin{subfigure}[b]{0.23\textwidth} 
    \includegraphics[width=0.9\linewidth]{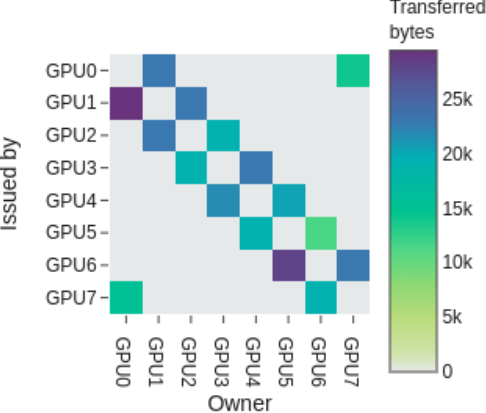}
    \end{subfigure}   
    \caption{a) Exact profiling, b) Sampling-based profiling. Figures are available under CC-BY \cite{snoopie_figures}.}
    \label{fig:sensitivity}
\end{figure}

Next, we evaluate the accuracy of \tool{} in capturing communication  when
monitoring the Stencil-p2p code with and without sampling.
Figure~\ref{fig:sensitivity} displays the captured communication matrix with the
exact profiling and when the sample size is 10\% of all memory operations. The
patterns are similar though the variation of color  indicates that the
communication volumes among  GPU pairs are less balanced in the sampling-based
profiling, which results from the stochastic nature of the sampling mechanism.

The forthcoming efforts will concentrate on improving the overhead of \tool{}
while preserving its accuracy. Additionally, we offer both exact and sampling
profiling modes, enabling programmers to choose between more detailed
examination of the code region using exact profiling or less accurate
sampling-based profiling.

\section{Related Work}

Table~\ref{tab:related-work} summarizes the comparison between \tool{} and the other communication profilers that work on GPUs. 
 Nsight Systems~\cite{nsight-system} and NVTAGS~\cite{nvtags} allow programmers to profile multi-GPU communications.  Tools like Eztrace~\cite{5948661}, Extrae\cite{extrae}, and Score-P\cite{10.1007/978-3-642-31476-6_7} generate traces of host-device and device-device data movements, however, all of these tools are limited to capture only explicit memory %
 and the latter three tracing tools support neither NCCL nor NVSHMEM. ComDetective captures inter-thread communication but only for multicore CPUs \cite{ComDetective}. ComScribe introduced in \cite{10.1007/978-3-030-71058-3_10} and extended in \cite{10.1007/978-3-031-06156-1_4}, captures P2P host-device, P2P device-device, and collective communications among multiple GPUs. 
 It %
 only presents the profiling results in the form of communication matrices. 

 Diogenes~\cite{10.1145/3295500.3356213} leverages the feed-forward measurement profiling model to identify problematic synchronization and data transfers through multi-stage/multi-run profiling of GPU applications. However, this tool does not capture inter-GPU remote memory operations.
 Li et al.~\cite{8573483} developed Tartan, which is a benchmark suite that characterizes scale-up and scale-out multi-GPU applications. Using this benchmark suite, they investigated the latency and bandwidth of various GPU bottlenecks. 
Another work by Pearson et al.~\cite{10.1145/3297663.3310299} introduced Comm$\vert$Scope, which is another benchmark suite that characterizes communications in multi-GPU machines. 

Paraver~\cite{paraver-paper} is a visualization tool for parallel programs running PVM message passing library~\cite{pvm3} that translates the trace files analyzed by DIMEMAS simulator~\cite{dimemas-paper} into graphical displays. 
VAMPIR~\cite{vampir-paper} is a graphical tool for MPI applications. It converts an input trace file into various graphical system views, which include %
the current activity of each processor, a timeline system view, and summary statistics of system behavior. PerfExplorer~\cite{1559993}  presents analysis results on parallel applications. It performs data mining on raw performance data stored in a PerfDMF DBMS~\cite{1488645} and presents the results in the forms of performance graphs, correlation scatterplots, and summary statistics. \texttt{EXTRAVIS}\cite{4268240} is a tool that visualizes execution traces in order to aid its users in comprehending the profiled applications. 

HPCToolkit\cite{hpctoolkit} presents its profiling results in a code-centric way using its graphical user interface, HPCViewer. It visualizes the source code of the profiled application and displays the number of events sampled using PMUs from each source code line. PerformanceHat~\cite{8811928} is another tool that presents the source code view of profiled applications in an IDE integration. It augments the view with monitoring data generated during the production run of the application and creates a performance model.
Kousha et al.~\cite{designing-a-profiling-and-visualization-tool-for} devised a visualization tool that leverages CUPTI~\cite{Cupti_Docs} and MPI\_T~\cite{MPI_T_Docs} to capture the utilization of GPU interconnects and correlates it with MPI communication patterns. 
Schaad et al.\cite{10.5555/3571885.3571970} developed a performance visualization tool that reports data movement and reuse behavior of profiled applications by leveraging static dataflow analysis. 
In contrast, our visualizer stands apart by offering fine-grained visualization of GPU communication. %

GVPROF~\cite{gvprof} and ValueExpert~\cite{valueexpert} use binary instrumentation for analyzing value-related inefficiencies in GPU kernels. DrGPUM~\cite{drgpum} employs binary instrumentation to detect inefficiencies related to memory usage. While these tools attribute identified inefficiencies to data objects and code lines, it's important to note that they are designed for single-GPU applications, unlike \tool{}.
\section{Conclusion}

Our work introduces \tool{}, a profiling tool that monitors and analyzes multi-GPU communication within a single node. \tool{} stands out from previous tools by being able to capture communication triggered by P2P direct accesses and NCCL/NVSHMEM library calls. To help locate performance bottlenecks and facilitate debugging, \tool{} supports communication attribution to the involved devices, source code lines, and data objects. The tool presents this information in a visualizer with support for multiple levels of granularity. We demonstrated its capabilities through various use cases. Future work will focus on increasing the tool's user-friendliness, adding support for multi-node communication, and applying it to a wider range of applications.

\begin{acks}
This work was supported by the European Research Council (ERC) under the European Union’s Horizon 2020 research and innovation programme under Grant 949587 and in part by the Royal Society-Newton Advanced Fellowship under Grant NAF\textbackslash{}R2\textbackslash{}202207.
\nocite{*}

\end{acks}

\bibliographystyle{ACM-Reference-Format}
\bibliography{references}

\end{document}